\documentclass[[linenumbers]{aastex63}
\usepackage{xcolor}
\usepackage{amsmath}
\usepackage{lipsum}
\usepackage{comment}
\usepackage{makecell}

\usepackage{hyperref}
\usepackage{csvsimple}

\usepackage{xcolor}            % colour text
\usepackage{ulem}                          % strikeout

\newcommand{\removed}[1]{\relax}\newcommand{\inserted}[1]{#1}
\newcommand{\skipthis}[1]{\relax}
\received{October 10, 2021}
\revised{January 12, 2022}
\accepted{January 28, 2022}
\submitjournal{The Astrophysical Journal Letters}

\shorttitle{Alpha Persei White Dwarfs}
\shortauthors{Miller et al.}
\graphicspath{{./}{figures/}}
\begin{document}

\title{The Ultramassive White Dwarfs of the Alpha Persei Cluster}

\correspondingauthor{David R. Miller}
\email{drmiller@phas.ubc.ca}

\author[0000-0002-4591-1903]{David R. Miller}
\affiliation{Department of Physics and Astronomy, University of British Columbia, Vancouver, BC V6T 1Z1, Canada}

\author[0000-0002-4770-5388]{Ilaria Caiazzo}
\affiliation{TAPIR, Walter Burke Institute for Theoretical Physics, Mail Code 350-17, Caltech, Pasadena, CA 91125, USA}

\author[0000-0001-9739-367X]{Jeremy Heyl}
\affiliation{Department of Physics and Astronomy, University of British Columbia, Vancouver, BC V6T 1Z1, Canada}

\author[0000-0001-9002-8178]{Harvey B. Richer}
\affiliation{Department of Physics and Astronomy, University of British Columbia, Vancouver, BC V6T 1Z1, Canada}

\author[0000-0001-9873-0121]{Pier-Emmanuel Tremblay}
\affiliation{Department of Physics, University of Warwick, Coventry, CV4 7AL, UK}

%\nocollaboration{4}



\begin{abstract}
We searched through the entire Gaia EDR3 candidate white dwarf catalogue for stars with proper motions and positions that are consistent with them  having escaped from the Alpha Persei cluster within the past 81~Myr, the age of the cluster. In this search we found five candidate white dwarf escapees from Alpha Persei and obtained spectra for all of them. We confirm that three are massive white dwarfs sufficiently young to have originated in the cluster. All these are more massive than any white dwarf previously associated with a cluster using Gaia astrometry, and possess some of the most massive progenitors.  In particular, the white dwarf Gaia~EDR3~4395978097863572, which lies within 25~pc of the cluster centre, has a mass of about 1.20 solar masses and evolved from an 8.5 solar-mass star, pushing the upper limit for white dwarf formation from a single massive star, while still leaving a substantial gap between the resulting white dwarf mass and the Chandrasekhar mass.

%The spectroscopy yields a temperature of $52,000\pm1,000$~K and $\log g=9.2 \pm 0.3$.  Combining the temperature measurement with the photometry gives a mass of 1.25~M$_\odot$ and an age of 26~Myr, meaning that the cluster produced a white dwarf at an age of 55~Myr from a 6.8~M$_\odot$ star.  
 
\end{abstract}

\keywords{stars: clusters -- massive -- supernovae -- white dwarfs -- Galaxy: open clusters}


\section{Introduction}

The maximum mass of a stable white dwarf (WD) has a widely accepted value of about $1.38~M_{\odot}$ \citep{1987ApJ...322..206N}; the maximum mass of a WD precursor star, however, is much more contentious. Theory suggests this value should be around $8~M_{\odot}$ \citep{1983A&A...121...77W}, but for this limit to hold, the observed type II supernovae (SNe) rate should be much higher \citep{2011ApJ...738..154H}. This dearth of observed type II SNe may point to a higher maximum mass, with some initial mass functions suggesting a maximum progenitor mass closer to $12~M_{\odot}$ \citep[e.g.][]{2003ApJ...598.1076K}. Better constraining this limit is important as it has a profound impact on a number of astrophysical quantities, including, but not limited to, the formation rates of compact objects and the metal enrichment rates of galaxies. 

To probe this limit we have been searching for massive WDs which are members of young open star clusters. Identifying massive WDs in young clusters is advantageous as it allows us to use the WD cooling age to estimate the progenitor mass as the main-sequence turnoff mass of the cluster at the time the WD was born. The breadth of modern stellar surveys greatly expands our ability to search for these objects; in particular, the precise parallaxes and proper motions measured by the Gaia survey \citep{2016A&A...595A...1G} allow us to select high-confidence cluster members using only astrometry and photometry. % particularly using the Milky Way stellar survey Gaia. 
%Unfortunately, despite the large amount of available data, 
Recently, a wide search for massive WDs in young clusters \citep{2021ApJ...912..165R} identified new young and high-mass WDs as cluster members, but failed to identify any cluster member WDs with masses in excess of $1.1~M_{\odot}$ or with progenitors over 6.2~M$_{\odot}$, leaving a gap in the high-mass region of the WD initial-final mass relation (IFMR). %\citep{2021ApJ...912..165R,pleiades}.

%have been unable to identify any cluster member WDs with masses in excess of $1.1~M_{\odot}$ or with progenitors over 6.2~M$_{\odot}$, leaving a gap in the high-mass region of the WD initial-final mass relation (IFMR) \citep{2021ApJ...912..165R,pleiades}. 

As the most massive WDs are the first to be born in a cluster, and since escape velocities in open clusters are quite small, the missing massive WDs might have escaped their host clusters. Open clusters are in fact known to be deficient in WDs \citep{2003ApJ...595L..53F}, and this deficit is thought to occur from WDs receiving a natal velocity kick of a few km/s at birth \citep[see][and references therein]{2009ApJ...695L..20F,2007MNRAS.382..915H}. In order to increase the number of potential massive WD cluster members, we expanded our search to include WDs that may have escaped from their host clusters.
%, as open clusters are known to be deficient in WDs \citep{2003ApJ...595L..53F}. This deficit is thought to occur from WDs receiving a natal velocity kick of a few km/s at birth \citep[see][and references therein]{2009ApJ...695L..20F,2007MNRAS.382..915H}. 
In previous work, we developed a technique that uses five-dimensional phase space information from Gaia EDR3 \citep{2020arXiv201201533G} to trace stars back to their potential birth clusters \citep{pleiades}. This technique was applied to a sample volume around each of the five nearest open clusters whose ages are less than 200 Myrs old \citep{youngclusters}. In the current paper, we expand the analysis for one of these young nearby clusters, Alpha Persei, and search the entire Gentile-Fusilo Gaia EDR3 WD catalogue \citep{2021arXiv210607669G} for candidate escapees. 

We describe the methodology used to identify escaped WDs in \S~\ref{sec:sample}, examine candidate massive WDs in \S~\ref{sec:white-dwarfs}, discuss their implication to the WD IFMR in \S~\ref{sec:IFMR}, and summarize our findings in \S~\ref{sec:conclusions}. In this search, we have identified five candidate massive WD escapees from the Alpha Persei cluster, two of which were also found in the aforementioned search in \citet{youngclusters}. We obtained follow-up spectroscopy for each of these five objects, and were able to confirm three of the WDs as high-confidence escapee massive WDs. These three are the most massive cluster WDs identified thus far. We estimate the most massive of these WDs to have a precursor whose mass is beyond the theoretical limit of 8~M$_{\odot}$. Given that the WD has a mass of about 1.2~M$_\odot$, notably below the Chandrasekhar limit of approximately 1.38~M$_{\odot}$, this finding hints at the idea that the upper limit on the mass of a star that can end its life as a WD is well above 8~M$_{\odot}$ or that single-star evolution does not produce WDs with masses all the way up to the Chandrasekhar limit.
%further supports the idea that the upper limit on the mass of a star that can end its life as a WD is well above 8~M$_{\odot}$.

\section{Sample}
\label{sec:sample}

Using Gaia EDR3 astrometry and radial velocities from Gaia DR2, \citet{youngclusters} calculated the mean velocity of the stars in the Alpha Persei cluster relative to the Sun, as well as the displacement of the centre of the cluster relative to the Sun as:
\begin{align}
    {\bf v}_\textrm{cluster} &= (-13.9\pm 0.8, -24.2 \pm 0.4,-6.83 \pm 0.2 )~\textrm{km/s} \\
    {\bf r}_\textrm{cluster} &= (-146.5\pm 0.7, 93.5 \pm 0.4,   -19.9 \pm 0.1)~\textrm{pc}
\end{align}
in Galactic coordinates.

%Using those stars within the sample that have radial velocities measured with Gaia DR2, we calculate the mean velocity of the stars within the Alpha Persei cluster relative to the Sun as 
%\begin{equation}
% {\bf v}_\textrm{cluster} = (-13.9\pm 0.8, -24.2 \pm 0.4,-6.83 \pm 0.2 )~\textrm{km/s}
% \label{eq:1}
 %\end{equation}
% in Galactic coordinates.  Using all of the stars within the cluster sample, we calculate the mean displacement of the cluster relative to the Sun as 
 % [-146.52115111   93.46317408  -19.85320989] [10.82077911  7.18465345  2.09859095] [0.66850997 0.44386937 0.12965138] 262
% \begin{equation}
% {\bf r}_\textrm{cluster} = (-146.5\pm 0.7, 93.5 \pm 0.4,   -19.9 \pm 0.1)~\textrm{pc},
% \label{eq:2}
% \end{equation}
% again in Galactic coordinates.  Both of these values agree within the uncertainties with those derived by \citet{2019A&A...628A..66L} from Gaia DR2 data.
 
To look for potential escapees from the Alpha Persei cluster, we determine the distance from the cluster of each object within the entire Gentile-Fusilo EDR3 WD catalogue \citep{2021arXiv210607669G} as a function of time, $d(t)$, assuming no relative acceleration and an arbitrary radial displacement ($\delta r$)
\begin{equation}
d(t)^2 = \left [ {\bf r}-{\bf r}_\textrm{cluster} 
+  t \left ({\bf v}_\textrm{2D}-{\bf v}_\textrm{cluster} \right ) + \hat{\bf{r}} \delta r  \right ]^2,
\label{eq:4}
\end{equation}
where ${\bf r}$ is the displacement of the star from the Sun, and ${\bf v}_\textrm{2D}$ is the velocity of the star in the plane of the sky. We then determine the time when the star and the cluster were or will be the closest together as
\begin{equation}
t_\textrm{min} = \frac{ \Delta {\bf r}\cdot  \Delta {\bf v} - \left (  \Delta {\bf r} \cdot \hat{\bf{r}}  \right ) \left (  \Delta {\bf v} \cdot \hat{\bf{r}}  \right ) }{\left (  \Delta {\bf v} \cdot \hat{\bf{r}}  \right )^2-\left (  \Delta {\bf v}\right )^2},
\label{eq:5}
\end{equation}
where
\begin{equation}
 \Delta {\bf r} = {\bf r}-{\bf r}_\textrm{cluster} ~\textrm{and}~
 \Delta {\bf v} = {\bf v}_\textrm{2D}-{\bf v}_\textrm{cluster}.
 \label{eq:6}
 \end{equation}
This also yields an estimate of the radial displacement and velocity of the star as
\begin{equation}
\delta r = v_r t_\textrm{min} =  -\hat{\bf{r}} \cdot \left ( \Delta {\bf r} +t_\textrm{min} \Delta {\bf v}   \right ) 
\label{eq:7}
\end{equation}
 and
 \begin{equation}
 \hat {\bf v}_\textrm{3D} = {\bf v}_\textrm{2D} + v_r \hat{\bf{r}}
 \label{eq:8}
 \end{equation}
 so
 \begin{equation}
 \Delta \hat {\bf v}_\textrm{3D} = \hat {\bf v}_\textrm{3D} - {\bf v}_\textrm{cluster}
 \label{eq:9}
 \end{equation}
 where the caret denotes that this is the reconstructed velocity.  To be deemed a candidate escapee, we take the distance of closest approach to be $d_\textrm{min}<15$~pc, and the time of closest approach to be during the lifetime of the cluster \citep{1999ApJ...510..266B,youngclusters}, $-81~\textrm{Myr} < t_\textrm{min}< 0$~Myr; also, we impose $|\Delta \hat {\bf v}_\textrm{3D}|<5$~km/s, determined by looking at the cumulative distribution of reconstructed relative 3D velocities of sample stars that met escapee criteria for $d_\textrm{min}$ and $t_\textrm{min}$ \citep{youngclusters}. Furthermore, to identify the potential WD escapees we further restrict the sample to WDs whose age is estimated to be less than 250~Myr and whose mass is greater than 0.85~M$_\odot$ from their Gaia EDR3 photometry, thus still allowing for the possibility that interstellar reddening and absorption could make the objects appear older and less massive than their true values.
 
\section{Candidate White Dwarf Escapees}
\label{sec:white-dwarfs}

\begin{figure*}
    \centering
    \includegraphics[width=\textwidth,clip,trim=0 3.39in 0 3.79in]{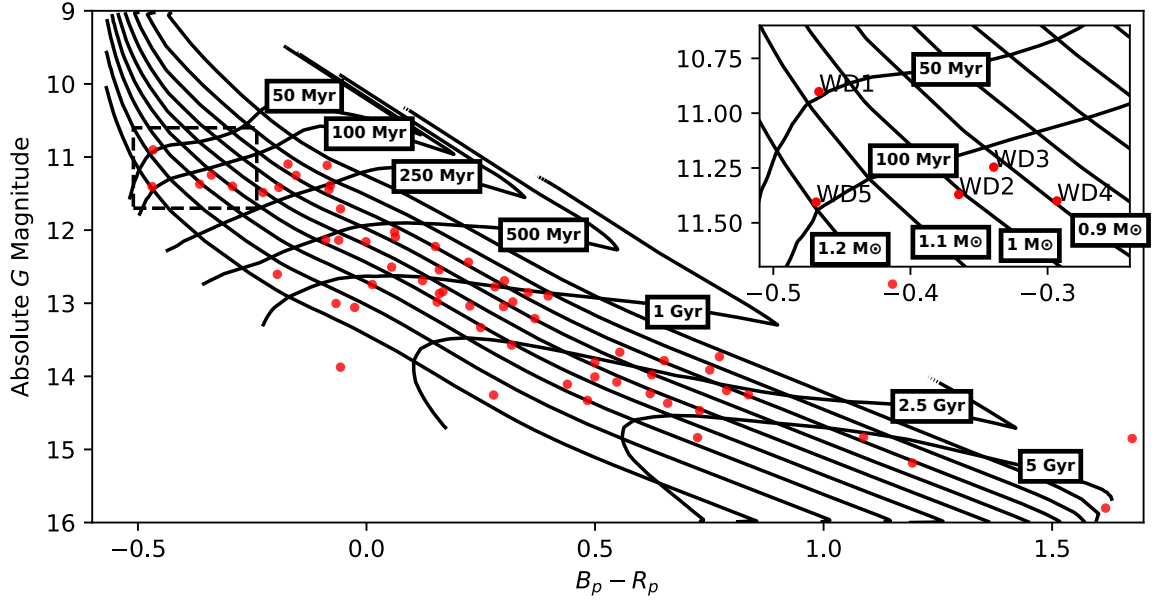}
    \caption{The candidate white dwarf escapees and members from the Alpha Persei cluster.  The effects of extinction have been removed using the mean value of $A_V$ from the \citet{2021arXiv210607669G} catalogue. The five white dwarfs that are both more massive than 0.85~M$_\odot$ and younger than 250~Myr (according the Gaia EDR3 data) are indicated in the cutout.  Contours of constant mass run from the top-left to the bottom-right with 0.4~M$_\odot$ at the top and increasing in increments of 0.1~M$_\odot$ to 1.2~M$_\odot$.  Contours of constant age run from the lower-left toward the upper-right with 50~Myr, 100~Myr, 250~Myr, 500~Myr, 1~Gyr, 2.5~Gyr and 5~Gyr from top to bottom. A reddening of $E(B-V)=0.1$ corresponds to $E(B_p-R_p)=0.15$, so the positions of the stars in this diagram depend sensitively on the assumed reddening; the masses inferred from photometry alone are likely to be imprecise.}
    \label{fig:cmd}
\end{figure*}

This search yielded five new candidates from the Alpha Persei cluster as shown in Fig.~\ref{fig:cmd}. Because of their current proximity to the cluster (within 25 pc) and small relative proper motions, \citet{2019A&A...628A..66L} have identified two of these white dwarfs (WD1 and WD2) as candidate members of the cluster. These two objects were were also identified in a search of the entire Gaia EDR3 database as candidate escapees of the cluster \citep{youngclusters}. The three others are now more than 100~pc away from the centre of the cluster. The astrometric, spectroscopic, and derived quantities for all five objects are presented in Tables~\ref{tab:wd_photo} and \ref{tab:wd_results}. 

\subsection{Spectroscopic Analysis}

We obtained optical spectroscopy for WD1 (Gaia EDR3 439597809786357248) and WD2 (Gaia EDR3 24400369345718860) with the 8.1m Gemini-North telescope using the Gemini Multi-Object Spectrograph \citep[GMOS;][]{2004PASP..116..425H,2016SPIE.9908E..2SG} in long-slit mode, using the B600 grating with no filter centered at 520~nm. Employing a 1.00 arcsecond focal plane mask, we binned 2x2 in both the spectral and spatial directions, providing a pixel scale of 0.161 arcsec/pixel, for a post binning resolution of $\approx 1$ angstrom. Total exposure time was 1~hr 3~mins for WD1, and 1~hr 19~mins for WD2. For WD5 (Gaia EDR3 1983126553936914816), we obtained spectra with the 10m Keck I Telescope (HI, USA) and the Low Resolution Imaging Spectrometer \citep[LRIS;][]{1995PASP..107..375O,1998SPIE.3355...81M}, using the R600 grism for the blue arm (R = 1100) and the R600 grating for the red arm (R = 1400) with 2x2 binning, for a total exposure time of 600 s. Spectra for WD3 (Gaia EDR3 1924074262608187648) and WD4 (Gaia EDR3 1990559596140812544) were obtained with the Double-Beam Spectrograph \citep[DBSP;][]{1982PASP...94..586O} on the 200 inch Hale telescope at Palomar observatory, with the R600 grating on the blue arm and the R316 grating on the red arm, for a total exposure time of 20 minutes each. 

The spectra, showing a blue continuum and broad hydrogen Balmer absorption lines, confirm that the five stars are WDs with hydrogen-dominated atmospheres (DA). The lack of notable Zeeman splitting of the spectral lines excludes the possibility of a strong magnetic field in  any candidate. 
%We fit the Balmer lines using Non-local Thermodynamic Equilibrium (NTLE) pure hydrogen hot WD models \citep{2011ApJ...730..128T}. The leftmost panel of Fig.~\ref{fig:WDall} displays the best fit model, revealing a hot WD with an effective temperature of nearly $42,000$~K and surface gravity of $10^{9.05}$~cm~s$^{-2}$. 
We analyse the spectroscopic data to obtain estimates for the surface gravities and temperatures of the five WDs. We employ atmospheric models developed by \cite{2010ApJ...720..581G} and by \cite{2011ApJ...730..128T}. In both sets of models, the hydrogen atmosphere is computed without the assumption of local thermodynamic equilibrium; the main difference is that in the former, the composition of the atmosphere includes carbon, nitrogen, and oxygen at solar abundance ratios, while the latter are made of pure hydrogen. The addition of metals in the atmosphere is important for very hot WDs, where metal levitation in the intense radiation field can change the shape of the Balmer lines \citep{2010ApJ...720..581G}.  Our fitting method to the Balmer lines is similar to the routine outlined in \citet{2005ApJS..156...47L}: we fit the spectrum with a grid of spectroscopic models combined with a polynomial in $\lambda$ (up to $\lambda^9$) to account for calibration errors in the continuum; we then normalize the spectrum using this smooth function picking normal points at a fixed distance in wavelength to the lines and finally use our grid of model spectra to fit the Balmer lines and extract the values of the effective temperature ($T_{\rm{eff}}$) and logarithm of the surface gravity ($\log \rm{g}$). The nonlinear least-squares minimization method of Levenberg-Marquardt is used in all our fits.

\begin{figure*}
\centering
  \setlength{\unitlength}{1cm}
  \begin{picture}(17,8)(-0.3,0)
 \put(0.53,0.725){\includegraphics[width=0.183\textwidth,clip,trim=1.335in 1.14in 0 0]{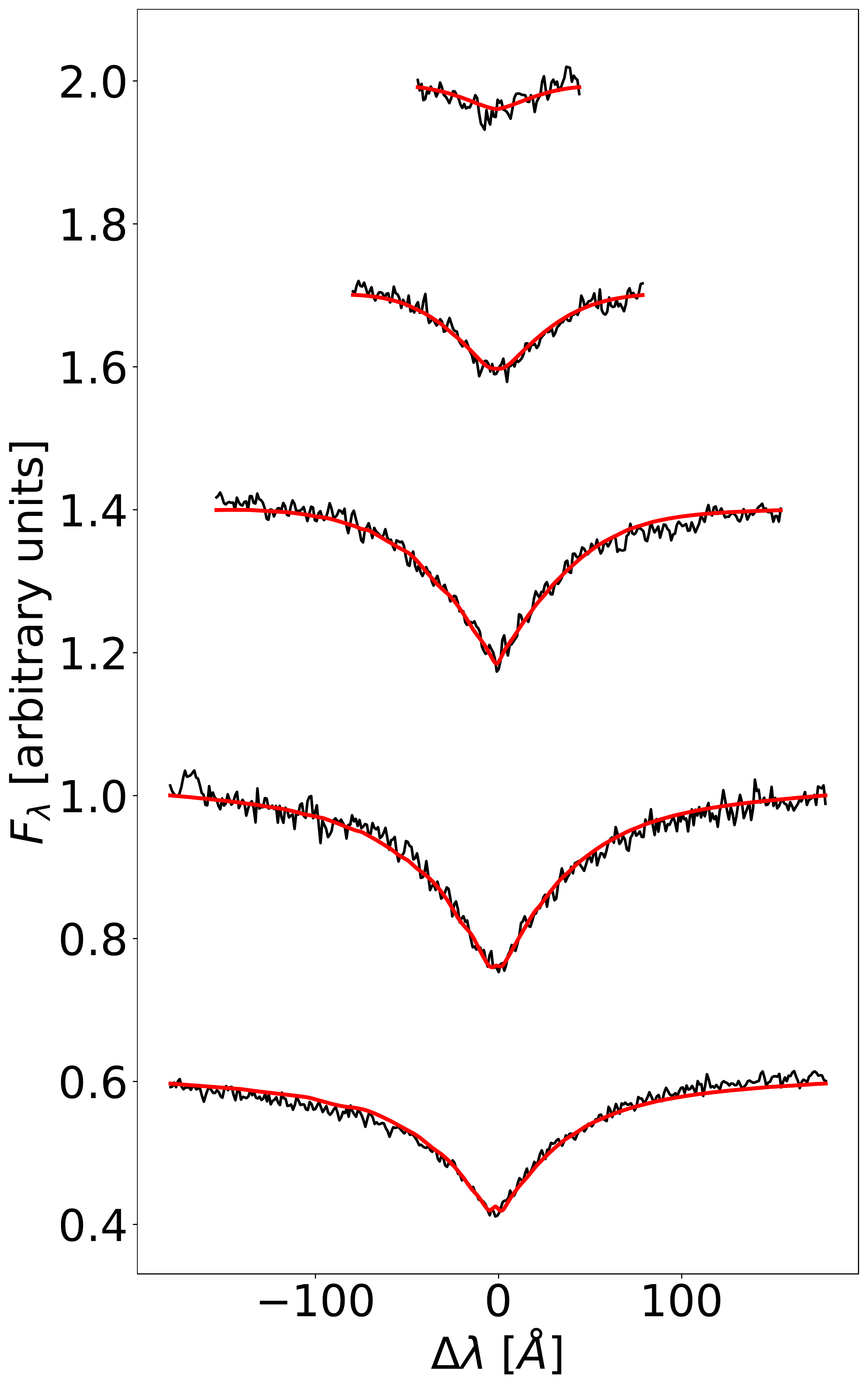}}
%\put(3.9,0.72){\includegraphics[width=0.18\textwidth,clip,trim=0.92in 0.75in 0 0]{figures/WD2_keck_polluted.pdf}}
\put(3.81,0.725){\includegraphics[width=0.183\textwidth,clip,trim=1.335in 1.14in 0 0]{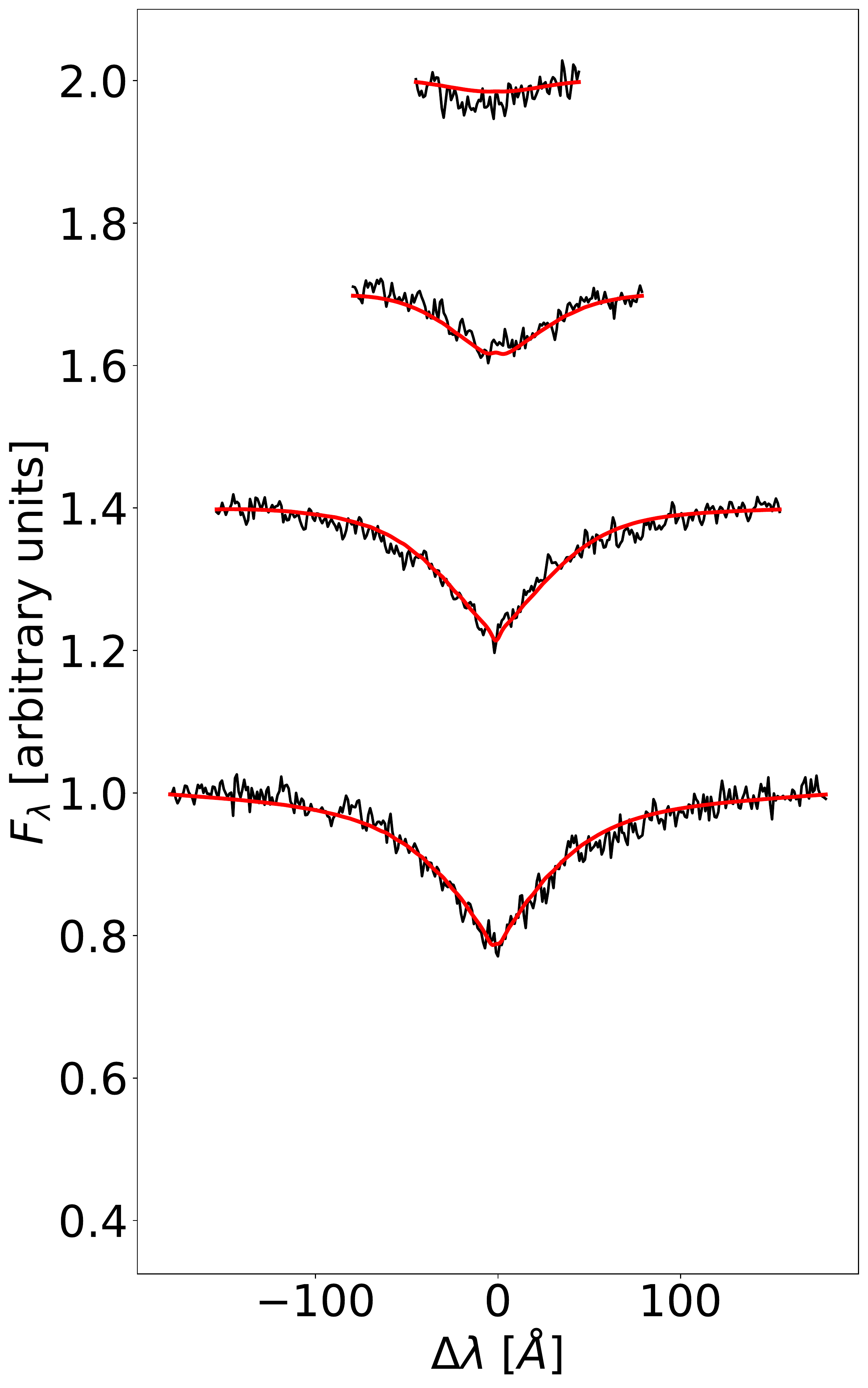}}
\put(7.05,0.72){
        \includegraphics[width=0.18\textwidth,clip,trim=0.92in 0.75in 0 0]{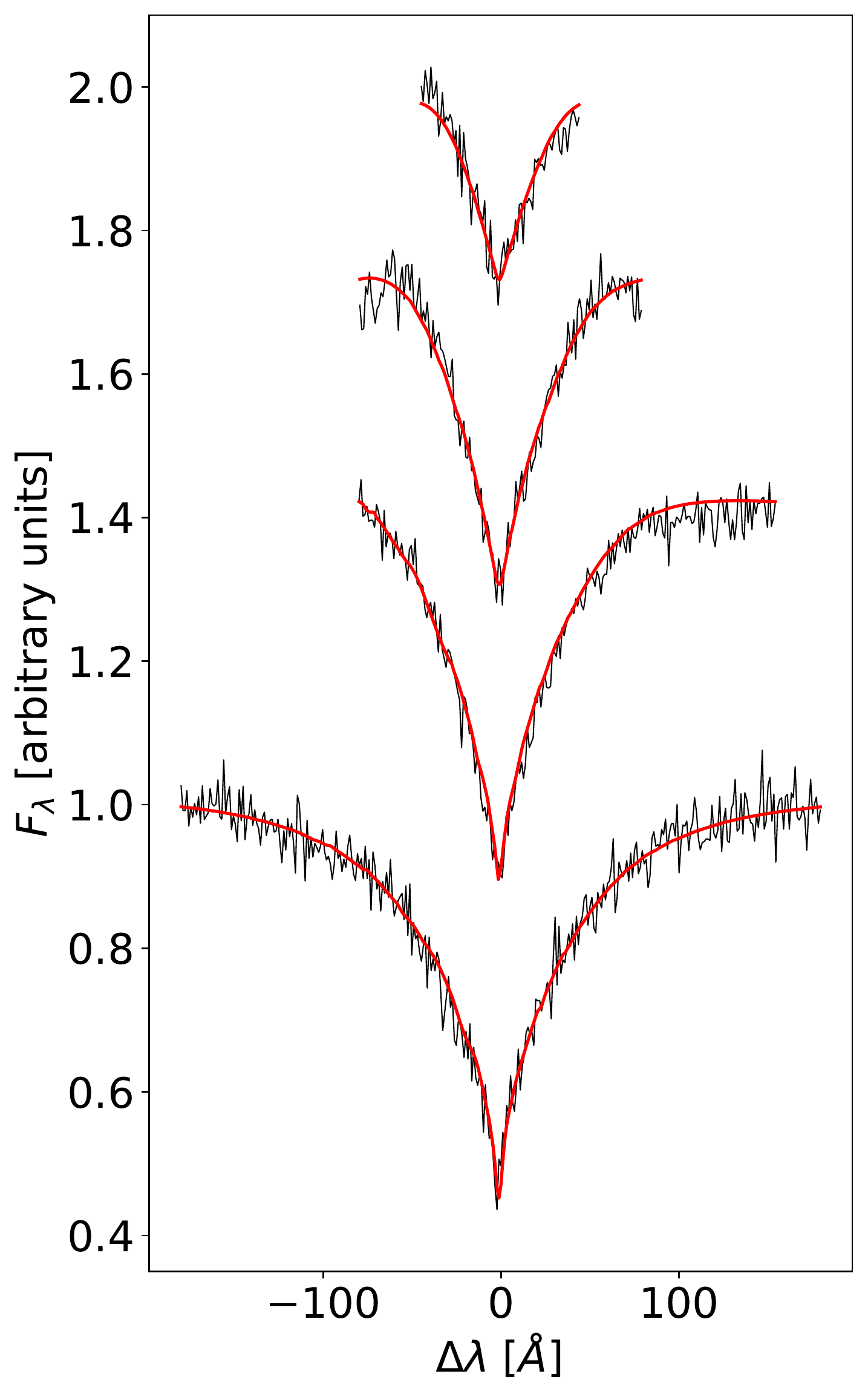}       }
\put(10.3,0.72){
        \includegraphics[width=0.18\textwidth,clip,trim=0.92in 0.75in 0 0]{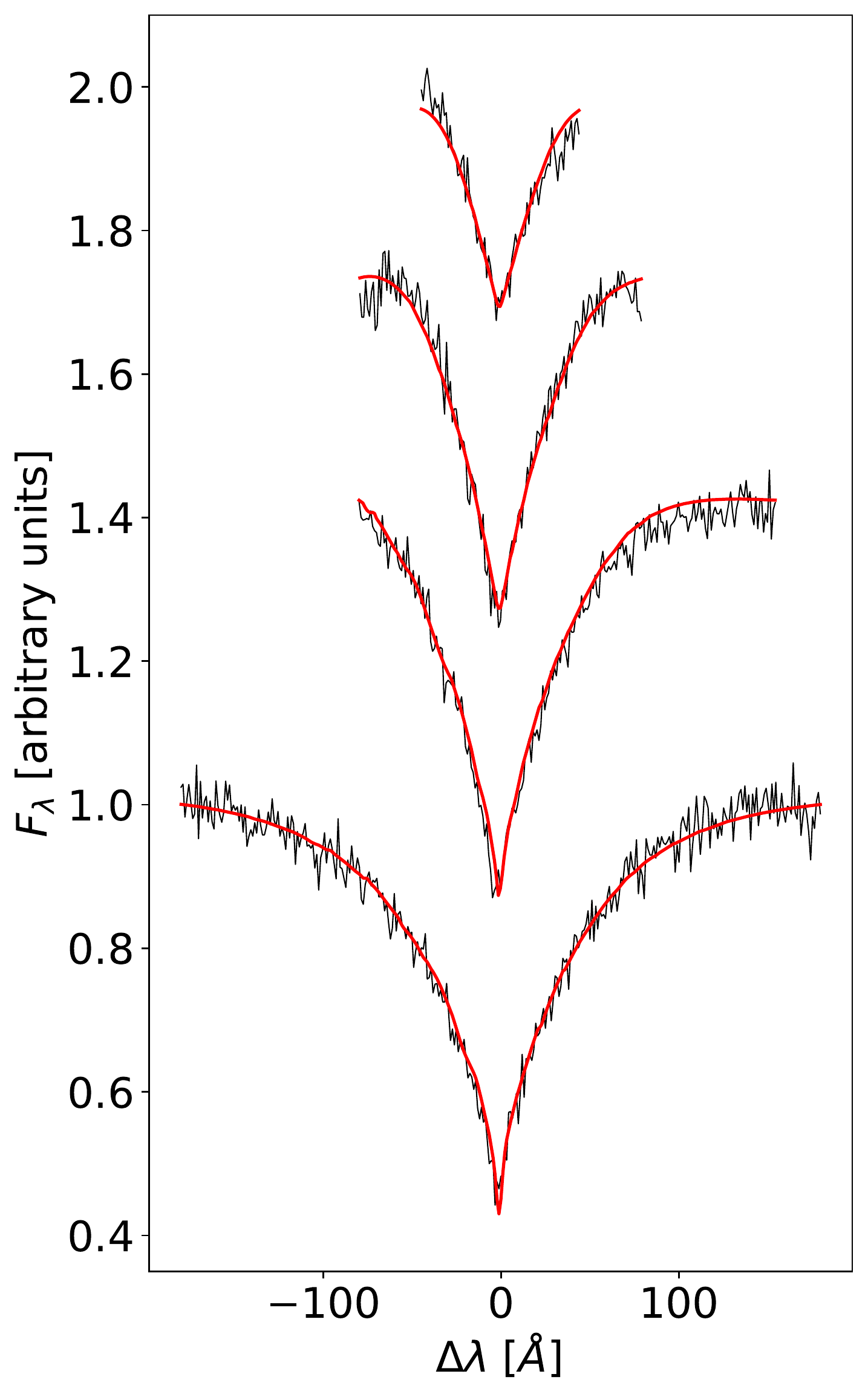}}
\put(13.55,0.72){
        \includegraphics[width=0.18\textwidth,clip,trim=0.92in 0.75in 0 0]{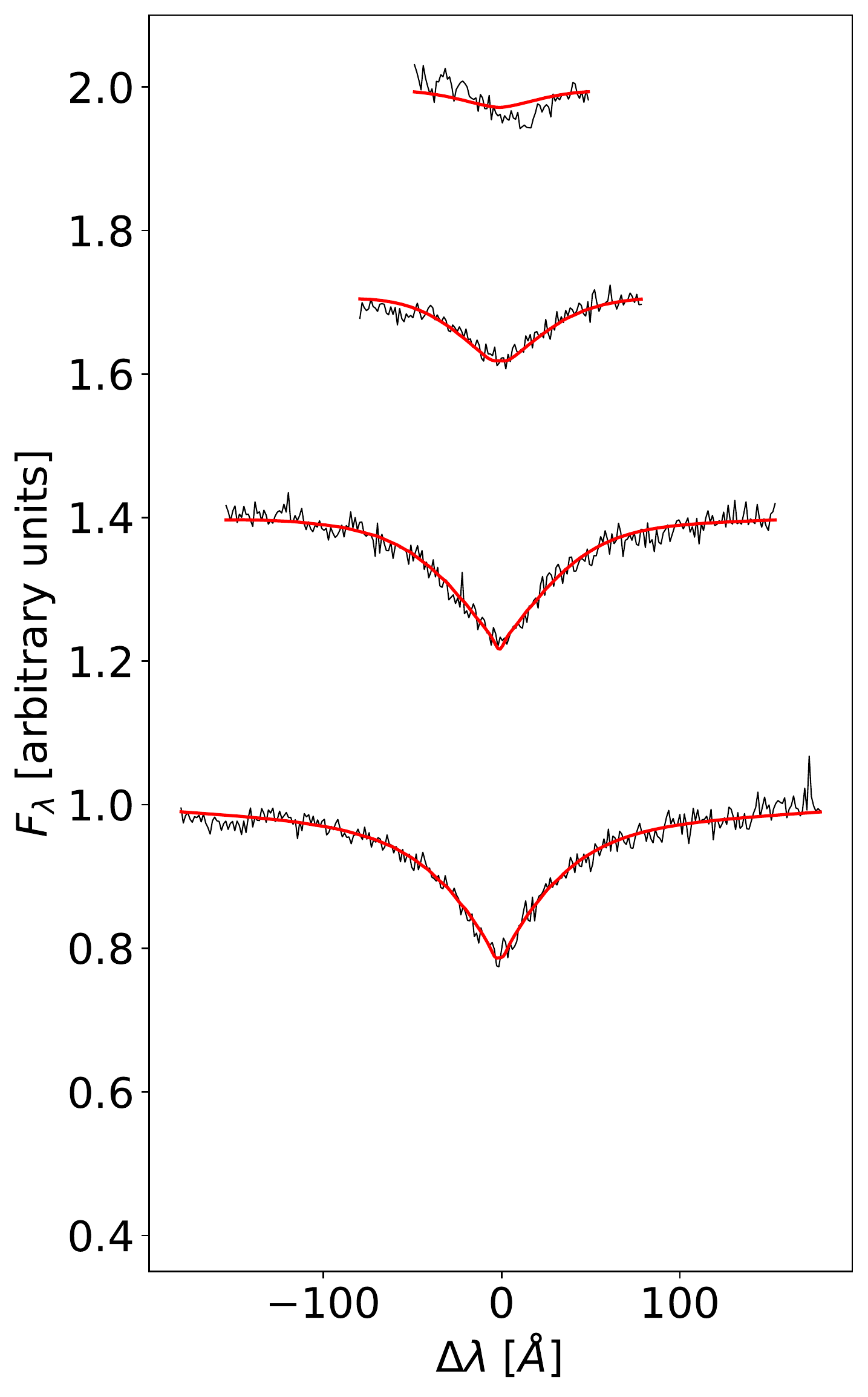}       }
\put(1.73,6.5){WD 1}
\put(5.035,6.5){WD 2}
\put(8.300,6.5){WD 3}
\put(11.555,6.5){WD 4}
\put(14.855,6.5){WD 5}      \put(2.125,0.4){$0$}
\put(0.9,0.4){$-100$}
\put(2.75,0.4){$100$}
\put(1.7,0){$\Delta \lambda$ [\AA]}
\put(5.43,0.4){$0$}
\put(4.205,0.4){$-100$}
\put(6.005,0.4){$100$}
\put(5.005,0){$\Delta \lambda$ [\AA]}
\put(8.695,0.4){$0$}
\put(7.470,0.4){$-100$}
\put(9.270,0.4){$100$}
\put(8.270,0){$\Delta \lambda$ [\AA]}
\put(11.95,0.4){$0$}
\put(10.725,0.4){$-100$}
\put(12.525,0.4){$100$}
\put(11.525,0){$\Delta \lambda$ [\AA]}
\put(15.20,0.4){$0$}
\put(13.975,0.4){$-100$}
\put(15.825,0.4){$100$}
\put(14.825,0){$\Delta \lambda$ [\AA]}
\put(-0.3,2){\rotatebox{90}{$F_\lambda$ [arbitrary units]}}
\put(0.1,0.85){0.4}
\put(0.1,1.5){0.6}
\put(0.1,2.13){0.8}
\put(0.1,2.75){1.0}
\put(0.1,3.35){1.2}
\put(0.1,4.0){1.4}
\put(0.1,4.65){1.6}
\put(0.1,5.3){1.8}
\put(0.1,5.95){2.0}
\end{picture}
    \caption{\textit{left to right}: Balmer series from H-$\alpha$ to H-$\epsilon$ (bottom to top) for WD1 (GMOS on Gemini North), and  H-$\beta$ (bottom) to H-$\epsilon$ (top) for WD2 (GMOS on Gemini North), WD3 and WD4 (Palomar DBSP), and WD5 (Keck LRIS) with the best-fitting hydrogen atmosphere models superimposed.}
    \label{fig:WDall}
\end{figure*}
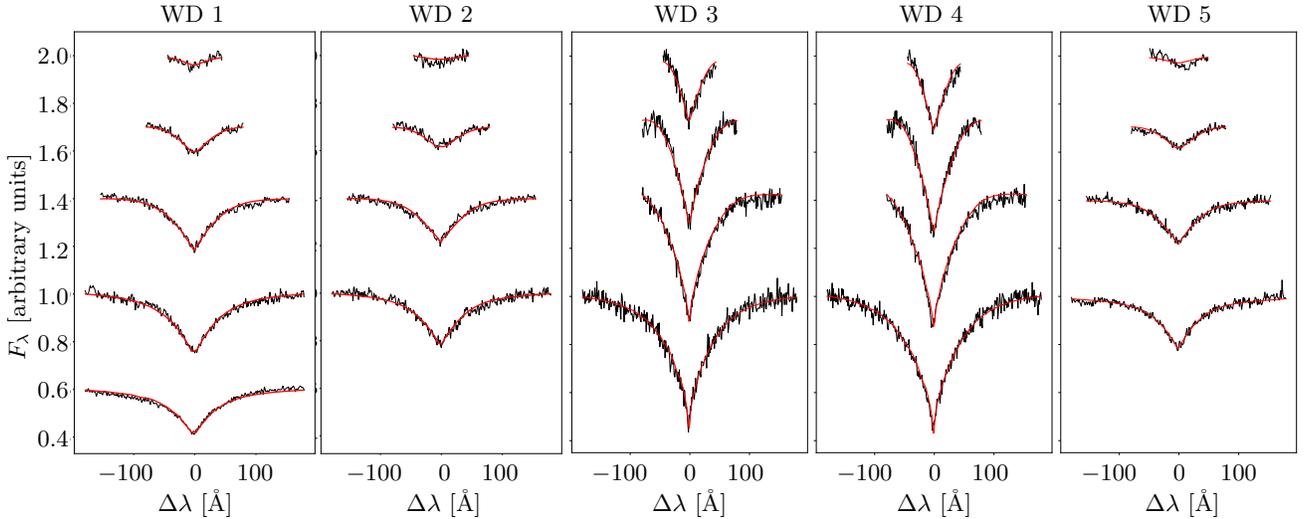

We initially fit each spectrum using the pure hydrogen atmosphere models of \citet{2011ApJ...730..128T}. As WDs 1, 2 and 5 all appear to be very hot WDs above $40,000$~K, we additionally fit using the \citet{2010ApJ...720..581G} models that include the influence of metals in the atmosphere. These models were developed because the presence of metals in the atmosphere, although not abundant enough to appear as additional metal lines, modify the shape of the Balmer lines, preventing simultaneous fitting of these lines. We find that WD1 does not display the Balmer line problem and the fit is not improved using metal-influenced models, while for WDs 2 and 5 the metal-influenced models notably improve the quality of the simultaneous fit to the Balmer lines. The small differences in the model fitting quality for these WDs is not surprising as the effective temperature of WD1 is approximately $42,000$~K, where the effects of metal levitation are supposedly very weak \citep{2010ApJ...720..581G}, while WDs 2 and 5 are at somewhat higher temperatures, closer to $T_{\rm{eff}} = 47,000$~K. The best fits are shown in Fig.~\ref{fig:WDall} using \citet{2011ApJ...730..128T} pure hydrogen atmosphere models for WDs 1, 3 and 4, and \citet{2010ApJ...720..581G} metal-influenced models for WDs 2 and 5. The resulting values of $T_{\rm{eff}}$ and $\log \rm{g}$ are listed in Table~\ref{tab:wd_results}. 

For each WD, we determine the mass and cooling age from two sets of high-mass WD cooling models: the \cite{2019A&A...625A..87C}  models, with an oxygen-neon (ONe) core composition and hydrogen-dominated atmosphere\footnote{http://evolgroup.fcaglp.unlp.edu.ar/TRACKS/ultramassive.html} as well as the \cite{2020ApJ...901...93B} thick hydrogen atmosphere models with a carbon-oxygen (CO) core\footnote{http://www.astro.umontreal.ca/$\sim$bergeron/CoolingModels/}. WDs should have a mass of at least $1.05$~M$_{\odot}$ to contain ONe cores \citep{2007A&A...476..893S}. In Table~\ref{tab:wd_results}, we list the masses and cooling ages of the five WDs, and, for WDs that have a mass above 1.05~M$_{\odot}$, we include the results of both the ONe and the CO fitting. Though we cannot strictly rule out the possibility of CO cores, theory suggests these three ultramassive WDs are all likely to have ONe cores. The CO core fit for WD1 implies a very massive precursor that would be a significant outlier in the IFMR, providing evidence that ONe is the preferred core composition for ultramassive massive WDs. Going forward, we will consider only the ONe results for these stars.

WDs 3 and 4 have masses close to 1~M$_{\odot}$ and cooling ages that are much longer than the age of the cluster. For this reason, they cannot be former members of Alpha Persei, and we remove them from our sample. WD1 escaped from the Alpha Persei cluster about 5~Myrs ago with a 3D escape velocity of 4.08~km/s. The ONe model fit suggests a very massive WD with a mass of $1.20\pm 0.01$~M$_{\odot}$ and a cooling age of $45\pm 4$~Myrs. Combined with the cluster age of $81\pm 6$~Myrs \citep{1999ApJ...510..266B,youngclusters}, this yields a precursor main sequence lifetime of $35\pm7 $~Myrs, corresponding to a $8.5\pm0.9~M_{\odot}$ progenitor according to the Padova isochrone models \citep{2012MNRAS.427..127B,2014MNRAS.445.4287T,2014MNRAS.444.2525C,2015MNRAS.452.1068C,2017ApJ...835...77M,2019MNRAS.485.5666P,2020MNRAS.498.3283P}. WD2 is currently about 20 pc away from the cluster, having escaped approximately 12 Myrs ago with a small escape velocity of 1.61~km/s. ONe models suggest a mass of $1.17\pm 0.01$~M$_{\odot}$ and a cooling age of $14\pm 4$~Myrs, giving a progenitor mass of $6.3\pm0.3~M_{\odot}$ from the Padova models. WD5 was possibly still a main sequence star when it escaped the cluster approximately 30 Myrs ago; a higher escape velocity of 4.43~km/s pushed the WD out to 136 pc away from the cluster. WD5 has a mass of $1.12\pm 0.02$~M$_{\odot}$ with a cooling age of $3\pm 1$~Myrs, corresponding to a $5.9\pm0.2~M_{\odot}$ progenitor.

\begin{table*}
\caption{Alpha Persei White Dwarf Candidates (Astrometric Quantities)}
\label{tab:wd_photo}
%\begin{tabular}{crrrrrrrrr}
\centering\begin{tabular}{crrrrrr}
\hline
       N & \multicolumn{1}{c}{Gaia EDR3 Source ID} &
       \multicolumn{1}{c}{RA}     &  \multicolumn{1}{c}{Dec}     & \multicolumn{1}{c}{Abs $G$}  &  \multicolumn{1}{c}{Parallax} & \multicolumn{1}{c}{$B_p-R_p$}           % & \multicolumn{1}{c}{$u_\textrm{AB}$} 
           \\
         & & \multicolumn{1}{c}{[deg]}  &  \multicolumn{1}{c}{[deg]}   &  \multicolumn{1}{c}{[mag]}   &   \multicolumn{1}{c}{[mas]}   & \multicolumn{1}{c}{[mag]}  \\
                            
\hline

1 & 439597809786357248 &  44.6805 & 50.3478 &  11.052  &  6.4358 &  $-0.379$     % & \multicolumn{1}{c}{---} 
\\
% Gaia DR2 244003693456233600
2 & 244003693457188608& 59.2417 &  45.0198 & 11.578 &  5.9335 & $-0.243$ % & \multicolumn{1}{c}{---} 
\\

3 & 1924074262608187648 & 354.1364 & 42.7338 & 11.321 & 6.6031 & $-0.295$  % & \multicolumn{1}{c}{---}
\\ %  4.68 112.92 -14.81 4.23    4.73 -23.30 0.92 22909
4 & 1990559596140812544 & 344.6288 & 53.7945 & 11.467 & 6.3103 & $-0.254$ % & \multicolumn{1}{c}{---} 
\\ %  4.49 116.88 -17.48 9.82    4.60 -24.77 0.89 20945
5 & 1983126553936914816 & 337.0616 & 45.5762 & 11.472 & 6.6416 & $-0.430$  % & \multicolumn{1}{c}{---}	
\\ %   4.29 136.21 -19.23 6.61    4.43 -30.00 1.21 35910
\end{tabular}
\end{table*}

%table removed to fit ApJ letters maximum tables+figures.

\begin{table*}
\caption{$\alpha$ Persei \inserted{White Dwarf} Candidates (Spectroscopic and Derived Quantities).  We use an age of $81\pm6$~Myr for the cluster \citep{youngclusters} and the Padova models to determine the initial masses.  The quantity $v_r$ is the reconstructed radial velocity of the star that brings it closest to the cluster centre in the past.}
\label{tab:wd_results}
\centering
\begin{tabular}{ccccccccccrcl}
\hline
N &
\multicolumn{1}{c}{$T_\textrm{eff}$} &
\multicolumn{1}{c}{$\log \rm{g}$} &
\multicolumn{1}{c}{$\Delta v_{2D}$} &  
\multicolumn{1}{c}{$d_\textrm{present}$} & 
\multicolumn{1}{c}{$v_r$} & 
\multicolumn{1}{c}{$d_\textrm{min}$} & 
\multicolumn{1}{c}{$\Delta v_\textrm{3D}$} &  
\multicolumn{1}{c}{$t_\textrm{escape}$} & 
Mass & 
$t_\textrm{cool}$ & 
Initial Mass &
\multicolumn{1}{c}{Comments} \\
  &
\multicolumn{1}{c}{[$10^3$ K]} & 
\multicolumn{1}{c}{[cm s$^{-2}$]} &
\multicolumn{1}{c}{[km/s]}   &  
\multicolumn{1}{c}{[pc]}                 & 
\multicolumn{1}{c}{[km/s]} & 
\multicolumn{1}{c}{[pc]} & 
[km/s] & 
[Myr] &
[M$_\odot$] & 
[Myr] & 
[M$_\odot$] &
 \\
\hline
1 &  $41.6(2)$ & $9.05(3)$ & 2.11 & ~24 & ~~$-6.1$  &  8.98 & 4.08 & ~5 & $1.20(1)$  &  $45(4)$  & $~~8.5(9)$  & ONe Core\\
  &            &           &      &     &           &       &      &    & $1.23(1)$  &  $62(4)$  & $12.^{+4}_{-2}~\,$   & CO Core\\
2 & $46.2(3)$  & $8.98(4)$ & 1.55 & ~20 & ~~~~$2.2$ &  5.40 & 1.61 & 12 & $1.17(1)$  &   $14(4)$  & $~~6.3(3)$    & ONe Core \\
  &            &           &      &     &           &       &      &    & $1.20(2)$  &  $31(4)$  & $~~7.2(6)$ & CO Core \\
%2 & $46.5(7)$  & $8.81(7)$ & 1.55 & ~20 & ~~~~$2.2$ &  5.40 & 1.61 & 12 & $1.10(3)$  &   $3(1)$  & $~~5.9(2)$    & ONe Core \\
%  &            &           &      &     &           &       &      &    & $1.13(3)$  &  $12(5)$  & $~~6.2(3)$ & CO Core \\
3 & $23.9(10)$  & $8.56(10)$ & 4.23 & 113 & $-14.8$   &  4.68 & 4.73 & 23 & $0.97(6)$  &  $133(11)$  & ---         & non-member\\
4 & $21.2(10)$  & $8.58(10)$ & 4.49 & 117 & $-17.5$   &  9.82 & 4.60 & 25 & $0.98(6)$  & $201(11)$  & ---         & non-member\\ 
5 & $47.5(5)$  & $8.84(5)$ & 4.29 & 136 & $-19.2$   &  6.61 & 4.43 & 30 & $1.12(2)$  &   $3(1)$  & $~~5.9(2)$ & ONe Core\\
  &            &           &      &     &           &       &      &    & $1.14(2)$  &  $12(4)$  & $~~6.2(3)$ & CO Core \\

\end{tabular}
\end{table*}

%\begin{figure*}
%    \centering
%        \includegraphics[width=0.213\textwidth,clip,trim=0 -0.15in 0 0]{figures/WD1.pdf}
%        \includegraphics[width=0.18\textwidth,clip,trim=0.92in 0 0 0]{figures/WD2_keck_polluted.pdf}
%        \includegraphics[width=0.18\textwidth,clip,trim=0.92in 0 0 0]{figures/WD3.pdf}       
%        \includegraphics[width=0.18\textwidth,clip,trim=0.92in 0 0 0]{figures/WD4.pdf}
 %       \includegraphics[width=0.18\textwidth,clip,trim=0.92in 0 0 0]{figures/WD5_keck_polluted.pdf}       
  %  \caption{\textit{left to right}: Balmer series from H-$\alpha$ to H-$\epsilon$ (bottom to top) for Gaia EDR3 439597809786357248 (GMOS on Gemini North), and  H-$\beta$ to H-$\epsilon$ (bottom to top) for Gaia EDR3 24400369345718860 (Keck LRIS), Gaia EDR3 1924074262608187648 and 1990559596140812544 (Palomar DBSP), and Gaia EDR3 198312655393691481 (Keck LRIS) with the best-fitting hydrogen atmosphere models superimposed.**This caption is confusing - can we make it more simple by numbering them?, then say (1) Balmer series from H-$\alpha$ (bottom) to H-$\epsilon$ (top) for Gaia EDR3 439597809786357248 (GMOS on Gemini North), H-$\beta$ to H-$\epsilon$ (bottom to top) for (2) Gaia EDR3 24400369345718860 (Keck LRIS), (3-4) Gaia EDR3 1924074262608187648 and 1990559596140812544 (Palomar DBSP) and (5) Gaia EDR3 198312655393691481 (Keck LRIS) with the best-fitting hydrogen atmosphere models superimposed.}
%    \label{fig:WDall}
%\end{figure*}

\subsection{The Possibility of Interlopers}

Because WD1 and WD2 were found in the well-defined sample of \citet{youngclusters}, we can estimate the number of young massive WDs that we would expect to appear in this sample by chance.   \citet{youngclusters} looked for stars in a volume of $6.4\times 10^6~\textrm{pc}^3$ surrounding the cluster.  Of the 463,917 stars in this volume, only 698 have kinematics consistent with having left the cluster within the last 100~Myr, and of these 698 stars about 300 may be interlopers \citep{youngclusters}.  On the other hand, \citet{massive} determined that there are 100 WDs with masses greater than 0.95~M$_\odot$ and ages less than 250~Myr within 200~pc of the Sun.  Combining these results we find that 0.012 young massive WDs would lie within the phase-space volume probed in the survey by chance. WD1 is both significantly younger and more massive than our thresholds, so the {\em a posteriori} chance of such a massive young WD being an interloper is a factor of fifteen smaller \citep{massive}, less than $10^{-3}$.  An alternative calculation that ignores the small relative proper motions between WD1 and WD2 and the cluster, but focuses on the small distance between them, is the number of massive young WDs that one would expect in an average sphere of 25~pc; the result is 0.04 WDs younger than 50~Myr and more massive than 0.95~M$_\odot$ \citep{massive}.

The remaining WDs, WD3 to WD5, were discovered in a broader search over the \citet{2021arXiv210607669G} catalogue, so we cannot assess the relative probabilities that WD5 comes from the cluster versus the field in the same manner as WD1 and WD2; however, in this case we can obtain an estimate by looking at all of the WDs that meet the kinematic criteria to have escaped from Alpha Persei (65 WDs).  The vast majority of these are too old to have originated from the cluster, and they are therefore clear interlopers.  \citet{massive} have found that only one out of one thousand WDs within 200~pc are more massive than 0.95~M$_\odot$ and younger than 250~Myr, yielding an expectation of finding only 0.06 WDs with these properties as chance interlopers within the sample.  Given this estimate, it is somewhat surprising that we did find two massive and relatively young interlopers in the sample, WD3 and WD4.  Their presence allows us to investigate the possibility that the phase-space volume corresponding to escapees from Alpha Persei contains a relative overabundance of massive stars and therefore massive WDs.   \citet{youngclusters} have argued that the Alpha Persei cluster lies on an orbit with a small inclination with respect to the Galactic plane.  If we assume that among these massive WDs (from 0.95 to 1.25~M$_\odot$) the usual relative distribution in mass and age occurs \citep{massive}, we can estimate the possibility that WD5 is also an interloper. The mean age of WD3 and WD4 is 167~Myr (fourteen times older than WD5) and they come from a population at least twice as large as that of WD5: WDs more massive than 0.95~M$_\odot$, compared to those more massive than 1.10~M$_\odot$.  This yields a probability for WD5 to be an interloper of 0.07, larger than the results for WD1 and WD2, but still very small. 

\section{The Initial-Final Mass Relation}   
\label{sec:IFMR}

\begin{figure}
    \centering
    \includegraphics[width=0.8\columnwidth]{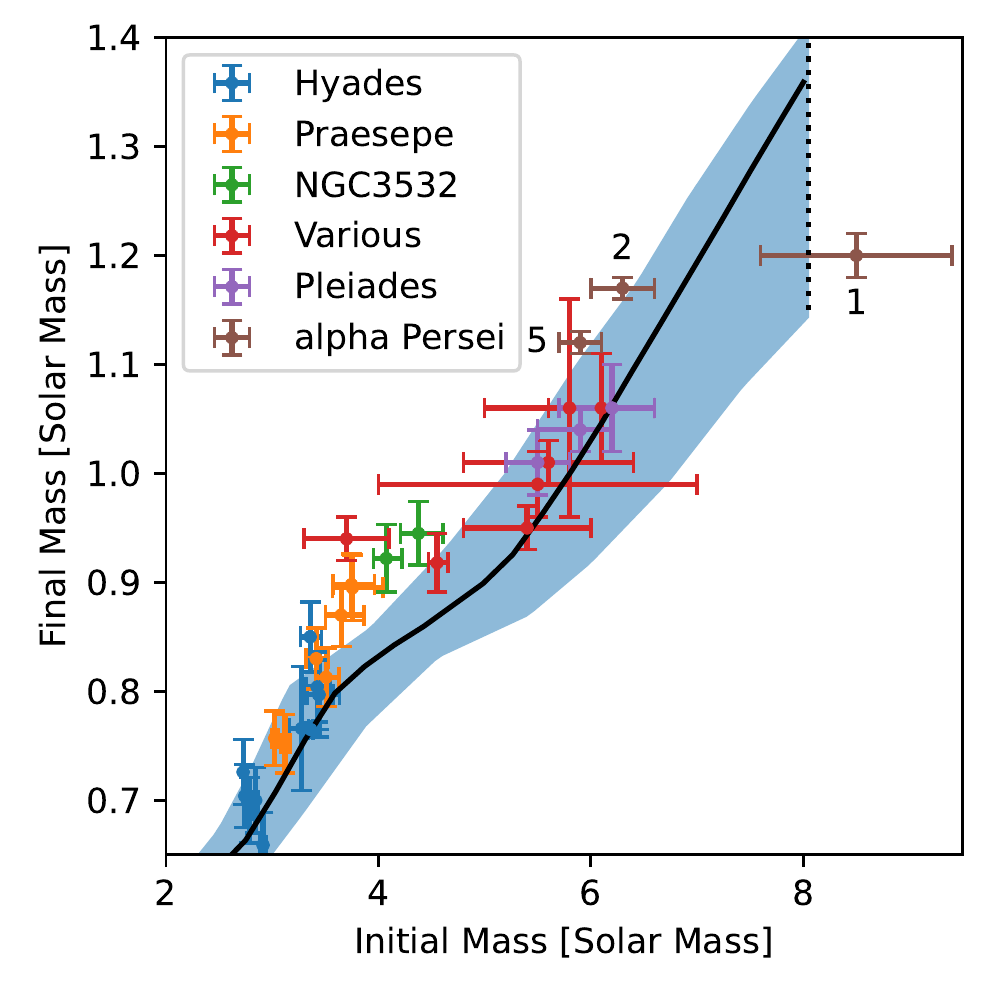}
    \caption{Initial-Final Mass Relation for White Dwarfs.  The ONe-model masses are plotted from Tab.~\ref{tab:wd_results}. Those labeled \inserted{``Pleiades'' are from \citet{pleiades}, those} from ``Various'' clusters are from \citet{2021ApJ...912..165R}, and the remainder are from \citet{2018ApJ...866...21C}.  \inserted{The black line and blue region denote the empirical initial-final-mass relation from \citet{2018ApJ...860L..17E} (up to 8~M$_\odot$) and its uncertainty bounds.}
    %The black curves trace the locus of possibilities for the five candidate escapees from Alpha Persei presented here. The curves trace the reddening values from  $E(B-V)=0.32$ down to the number indicated.  The unlabeled curves from upper and lower correspond to white dwarfs two and three with minimum reddenings of 0.17 and 0.10, respectively.
    }
    \label{fig:aper-IFMR}
\end{figure}

\inserted{Fig.~\ref{fig:aper-IFMR} shows the updated IFMR from \citet{2021ApJ...912..165R} including these new escaped cluster member WDs as well as those identified in \citet{pleiades}. Each of the three newly identified WDs from this work are more massive than any cluster WDs previously identified. For each of these new WDs we display results of ONe core fits. WD1 has a precursor mass of $8.5\pm0.9$~M$_{\odot}$, placing it near the theoretical limit of about $8$~M$_{\odot}$ \citep[e.g.][]{RevModPhys.74.1015}. Given that the WD's mass is still well below the Chandrasekhar mass, this supports an increased main-sequence upper mass limit for WD production, more consistent with expectations from observed SN II rates, or hints to the fact that single-star evolution does not produce WDs with masses all the way up to the Chandrasekhar limit.} 

\inserted{While field stars do not provide the initial mass of a star, cluster membership is required for that, it is nevertheless instructive to inquire as to the maximum mass of WDs seen outside of clusters. In an analysis of the 100 pc sample from the Montreal White Dwarf Database, \cite{2021MNRAS.503.5397K} identified 25 WDs with masses above 1.3~M$_{\odot}$ if all possess H-atmospheres and CO cores. If the WDs instead have ONe cores, which we expect is likely the case for the majority of WDs in this mass range, just 2 of them have masses above 1.3~M$_{\odot}$. However, 23 of the 25 would have masses above 1.25~M$_{\odot}$, well above the most massive WD in the current IFMR. Note, at a minimum, a third of these are merger remnants as revealed by high magnetic fields and rapid rotation. Nevertheless, their findings suggest that WDs are formed well closer to the Chandrasekhar limit than those that have been thus far identified in clusters, which when coupled with our results provides additional support for an increased upper limit on WD production.}

\section{Conclusions}   
\label{sec:conclusions}

We employ a technique that we developed in \citet{pleiades} for identifying WDs that may have escaped from open star clusters. In \citet{pleiades} as well as the follow up paper \citet{youngclusters}, this technique was used for a sample volume around five nearby young clusters. Here, we instead search the entire Gentile-Fusilo Gaia EDR3 WD catalogue for one specific cluster, Alpha Persei. By tracing the historical position of these catalogue WDs, we identified five candidates whose motion suggested they may have escaped from the cluster. Each of these were followed up with spectroscopy from   Gemini-North GMOS, Keck LRIS, or Palomar DBSP. 

The surface gravity and temperature of each WD was determined from the best fit NTLE hydrogen atmosphere models. From these results the mass and cooling age were determined using CO core cooling models as well as ONe models for those above $1.05~M_{\odot}$. Of the five WDs, three are consistent with being escaped former cluster members, while the other two have cooling ages which are larger than the age of the cluster, thus eliminating the possibility of membership. The progenitor mass for each of the three escaped members were determined using Padova isochrone models \citep{2012MNRAS.427..127B,2014MNRAS.445.4287T,2014MNRAS.444.2525C,2015MNRAS.452.1068C,2017ApJ...835...77M,2019MNRAS.485.5666P,2020MNRAS.498.3283P}. 

Though the results of this work provide valuable insight into the WD IFMR, we have not yet identified any cluster member WDs near the Chandrasekhar limit. Measurement uncertainty likely limits the technique we have used here to a handful of nearby clusters. That said, given that we have identified a significant number of escaped WDs, we are led to believe that many of these objects have also escaped from other clusters and are merely waiting to be identified. We will continue to work to develop techniques to identify these escaped WDs in the future to better constrain the upper mass limit of WD progenitor stars. 

\section*{Acknowledgements}

This work was supported in part by NSERC Canada and Compute Canada. I.C. is a Sherman Fairchild Fellow at Caltech and thanks the Burke Institute at Caltech for supporting her research.

This research has made use of the SIMBAD and Vizier databases, operated at CDS, Strasbourg, France and the Montreal White Dwarf Database produced and maintained by Prof. Patrick Dufour (Universit\'e de Montr\`eal) and Dr. Simon Blouin (LANL), 

% This research has made use of data obtained from or tools provided by the portal exoplanet.eu of The Extrasolar Planets Encyclopaedia.

This work has made use of data from the European Space Agency (ESA) mission
{\it Gaia} (\url{https://www.cosmos.esa.int/gaia}), processed by the {\it Gaia}
Data Processing and Analysis Consortium (DPAC,
\url{https://www.cosmos.esa.int/web/gaia/dpac/consortium}). Funding for the DPAC
has been provided by national institutions, in particular the institutions
participating in the {\it Gaia} Multilateral Agreement.

This work includes results based on observations obtained at the international Gemini Observatory, a program of NSF’s NOIRLab, which is managed by the Association of Universities for Research in Astronomy (AURA) under a cooperative agreement with the National Science Foundation. on behalf of the Gemini Observatory partnership: the National Science Foundation (United States), National Research Council (Canada), Agencia Nacional de Investigaci\'{o}n y Desarrollo (Chile), Ministerio de Ciencia, Tecnolog\'{i}a e Innovaci\'{o}n (Argentina), Minist\'{e}rio da Ci\^{e}ncia, Tecnologia, Inova\c{c}\~{o}es e Comunica\c{c}\~{o}es (Brazil), and Korea Astronomy and Space Science Institute (Republic of Korea).

Some of the data presented herein were obtained at the W. M. Keck Observatory, which is operated as a scientific partnership among the California Institute of Technology, the University of California and the National Aeronautics and Space Administration. The Observatory was made possible by the generous financial support of the W. M. Keck Foundation. 

% This publication makes use of data products from the Wide-field Infrared Survey Explorer, which is a joint project of the University of California, Los Angeles, and the Jet Propulsion Laboratory/California Institute of Technology, funded by the National Aeronautics and Space Administration. 

%Based on data products from observations made with ESO Telescopes at the La Silla Paranal Observatory under programme ID 177.D-3023, as part of the VST Photometric H{alpha} Survey of the Southern Galactic Plane and Bulge (VPHAS+, www.vphas.eu).

The Pan-STARRS1 Surveys (PS1) and the PS1 public science archive have been made possible through contributions by the Institute for Astronomy, the University of Hawaii, the Pan-STARRS Project Office, the Max-Planck Society and its participating institutes, the Max Planck Institute for Astronomy, Heidelberg and the Max Planck Institute for Extraterrestrial Physics, Garching, The Johns Hopkins University, Durham University, the University of Edinburgh, the Queen's University Belfast, the Harvard-Smithsonian Center for Astrophysics, the Las Cumbres Observatory Global Telescope Network Incorporated, the National Central University of Taiwan, the Space Telescope Science Institute, the National Aeronautics and Space Administration under Grant No. NNX08AR22G issued through the Planetary Science Division of the NASA Science Mission Directorate, the National Science Foundation Grant No. AST-1238877, the University of Maryland, Eotvos Lorand University (ELTE), the Los Alamos National Laboratory, and the Gordon and Betty Moore Foundation. 

Gemini spectra were processed using the Gemini IRAF package. LRIS spectra were reduced using the Lpipe pipeline \citep{2019PASP..131h4503P}. DBSP Spectra were reduced using the DBSP\_DRP pipeline \citep{2021arXiv210712339R}.

\software{Astropy \citep{astropy:2013, astropy:2018}}, WD$\_$models (\url{https://github.com/SihaoCheng/WD_models}).

\facilities{Gaia (DR2 \& EDR3), Gemini-North (GMOS), Keck:I (LRIS), Palomar (DBSP).}

%%%%%%%%%%%%%%%%%%%%%%%%%%%%%%%%%%%%%%%%%%%%%%%%%%
\section*{Data Availability}

We constructed the cluster escapee white dwarf catalogue from \inserted{the Gentile-Fusilo Gaia EDR3 WD catalogue available at} \url{https://warwick.ac.uk/fac/sci/physics/research/astro/research/catalogues/gaiaedr3_wd_main.fits.gz} \inserted{Data from GALEX and Pan-STARRS1 were obtained with Vizier and used in preliminary analysis}.



\bibliographystyle{aasjournal}
\bibliography{main} % if your bibtex file is called example.bib

%\begin{thebibliography}{99}
%\end{thebibliography}

%%%%%%%%%%%%%%%%%%%%%%%%%%%%%%%%%%%%%%%%%%%%%%%%%%

%%%%%%%%%%%%%%%%% APPENDICES %%%%%%%%%%%%%%%%%%%%%

\label{lastpage}

\end{document}